\documentclass[singlecolumn,superscriptaddress,amssymb,amsmath,nobibnotes,aps,
prd,showpacs,nofootinbib]{revtex4}%
\usepackage{graphicx}
\usepackage{epsf}
\usepackage{bm}
\usepackage{amsmath}
\usepackage{amsfonts}
\usepackage{amssymb}
\usepackage{epstopdf}
\usepackage{natbib}%
\usepackage{rotating}
\usepackage{pdflscape}
\providecommand{\U}[1]{\protect\rule{.1in}{.1in}}
\newcommand{\be}{\begin{equation}}
\newcommand{\ee}{\end{equation}}

\newcommand{\mincir}{\raise
-3.truept\hbox{\rlap{\hbox{$\sim$}}\raise4.truept\hbox{$<$}\ }}
\newcommand{\magcir}{\raise
-3.truept\hbox{\rlap{\hbox{$\sim$}}\raise4.truept\hbox{$>$}\ }}

\usepackage{color}

\begin{document}

\title{Gravitational baryogenesis in running vacuum models}

\author{
V.~K.~Oikonomou}
\email{v.k.oikonomou1979@gmail.com}
\affiliation{
 Tomsk State Pedagogical University, 634061 Tomsk, Russia\\}
\affiliation{Laboratory for Theoretical Cosmology, Tomsk State University of Control Systems
and Radioelectronics (TUSUR), 634050 Tomsk, Russia\\}

\author{Supriya Pan}
\email{span@iiserkol.ac.in}
\affiliation{Department of Physical Sciences, Indian Institute of Science Education and
Research$-$Kolkata, Mohanpur$-$741246, West Bengal, India}

\author{Rafael C. Nunes}
\email{rafadcnunes@gmail.com}
\affiliation{Departamento de F\'isica, Universidade Federal de Juiz de Fora, 36036-330, Juiz de Fora, MG, Brazil}

\pacs{98.80.-k, 95.36.+x, 95.35.+d}

\begin{abstract}
We study the gravitational baryogenesis mechanism for
generating baryon asymmetry in the context of running vacuum models.
Regardless if these models can produce a viable
cosmological evolution, we demonstrate that they
produce a non-zero baryon-to-entropy ratio
even if the universe is filled with conformal matter.
This is a sound difference between the running
vacuum gravitational baryogenesis
and the Einstein-Hilbert one, since in the latter
case, the predicted baryon-to-entropy ratio is zero.
We consider two well known and most used running
vacuum models and show that the resulting
baryon-to-entropy ratio is compatible with
the observational data. Moreover, we also show
that the mechanism of gravitational baryogenesis
may constrain the running vacuum models.
\end{abstract}
\maketitle

\section{Introduction}
\label{sec:intro}

The dynamics of the universe is yet a mysterious and unpredictable
chapter in modern physics. The addition of late time accelerated
expansion of the universe as a new chapter to its dynamical history
has been the central theme for the last couple of years. To account
such late accelerated expansion, usually one adds a new degrees of
freedom associated with an exotic component called dark energy
\cite{de1} in the framework of general relativity \footnote{However,
it deserves to be mentioned that the theory of gravitational
modifications has also been found to explain the late time
accelerated expansion without any need of dark energy
\cite{reviews1.1,reviews1.2,reviews1.3,reviews2,reviews3,%
reviews4,reviews5,sergoik,oik1,Cai:2015emx}.}.
The simplest dark energy candidate that accommodates the
astronomical data quite well is the $\Lambda$-cosmology, where
$\Lambda> 0$, is the time independent constant.  However,  the
$\Lambda$ term which is equivalent to the vacuum energy density,
$\rho_{\Lambda}= \Lambda/8 \pi G$ ($G$ is the Newton's gravitational
constant), leads to the cosmological constant problem that has remain
been a serious issue in modern cosmology \cite{weinberg}. Motivated
by the cosmological constant problem, a series of alternative
cosmological theories have been proposed in the literature, see
\cite{de1}, that gained a massive interest in the scientific
community so far. Amongst a class of distinct cosmologies, in this
work we are interested to discuss the models where the vacuum energy
is dynamical, i.e. when $\rho_{\Lambda}= \rho_{\Lambda} (t)$. In
other words, when the cosmological constant $\Lambda$, is replaced
by a time dependent term $\Lambda (t)$. From the phenomenological
ground this kind of models have been accepted as consistent
cosmological theories and in the last couple of years are very well
explored in detail \cite{rnv0,rnv0.1, rnv0.2, rnv1,rnv2,rnv3,rnv3.1,
rnv4a,rnv4b, rnv4c, rnv5,rnv6, rnv6.1,rnv7,rnv8,rnv9,rg2,Guberina,Shapiro}. In such
models the equation of state of the vacuum is strictly equal
to `$-1$', but the energy density of vacuum is variable throughout
the history of the universe. However, here we shall focus on a  more
particular class of dynamical vacuum models, known as the running
vacuum models (RVMs).
\\

The RVM models in agreement with the observational data
(\cite{Javier} and references there) have the capability to describe
the dynamical histroy of the universe in an effective way. In fact,
the RVM models are quite good in compared to some other
phenomenological models as discussed in Refs. \cite{Grande:2011xf,Lima:2012mu, 
Perico:2013mna, BLS2013, Basilakos:2013vya}. Further, it is also interesting to note that
this class of models carry an explanation to the time evolution of
the fundamental constants of nature, see \cite{constant1,constant2}
and also Ref. \cite{Sola2013,SV2015} for a review on RVMs. We remark
that very recently it has been argued that current observational
data might favor the dynamical nature of the vacuum
\cite{rnv10,rnv11,rnv12,rnv13,coupled4, coupled5, coupled6}. The
motivation of the current work is not to investigate whether the
RVMs are the alternative description to some other existing cosmological
theories, but to investigate the feasibility of the gravitational
baryogenesis mechanism in such models.\\

It is well known that baryon  asymmetry is one of the most
persisting problems in modern theoretical physics and cosmology, and
it refers to the observed excess of matter over anti-matter. The
existing observational data, mainly coming from the cosmic microwave
background \cite{bb2} in conjunction with the successful predictions
of Big Bang nucleosynthesis, indicate that matter overwhelms
anti-matter. Towards this research line, the gravitational
baryogenesis mechanism was proposed some time ago in Ref. \cite{gb1}
in order to explain the excess of matter over anti-matter in the
observable universe. Ever since, many modifications or
generalizations of the original gravitational baryogenesis
mechanism, have been proposed in the literature, see for example
\cite{gb1,gb2,gb3,gb4,gb5,gb6,gb7,gb8,gb9,gb10,gb11,gb12,gb13}. In
this paper we adopt the approach of the original gravitational
baryogenesis mechanism proposed in \cite{gb1}, and we investigate
what is the predicted baryon-to-entropy ratio in the context of
running vacuum theories. Thus, we considered two well known and most used running vacuum
models, the essential features of which we presented in some detail
in the next section. For these models we have calculated the
baryon-to-entropy ratio and investigated under which conditions  the
predicted baryon-to-entropy ratio can be compatible with the
observed value. The most interesting and the important finding in
this work is that the baryon-to-entropy ratio produced by the
gravitational baryogenesis mechanism in the presence of a variable
vacuum energy is non-zero even in the case that relativistic
massless matter is the only matter ingredient of the universe, with
regard to perfect fluids. This is different from the
Einstein-Hilbert gravity case, where in the presence of radiation
matter fluids, the predicted baryon-to-entropy ratio coming from
gravitational baryogenesis mechanism is zero, as was showed in
\cite{gb1}. In addition to this result, we also discuss that  the
compatibility
of the models  with observations can be achieved. In fact, we find that the observed baryon-to-entropy ratio may be used to constrain the class of RVMs. \\

This paper is organized as follows: In Section \ref{Field-equations} we present the theoretical framework of  running vacuum theory and  describe the models we shall use. Also, we fix the geometric conventions and explain some feature of running vacuum theories, essential to the later sections. In Section \ref{gravibarmain} we briefly discuss the gravitational baryogenesis mechanism where we assume that it generates the baryon asymmetry and calculate the predicted baryon-to-entropy ratio for the running vacuum models of
the previous section. In addition, we present some general considerations regarding the predicted baryon-to-entropy
ratio in the case that the only perfect fluid present is radiation. Moreover, we compare the results of the running
vacuum theory with the Einstein-Hilbert case and  investigate in which cases the results are compatible  with observations.
Finally, the conclusions containing our findings follow at the end of the paper.

\section{Dynamical Vacuum models}

\label{Field-equations}

In the present work we are interested in investigating the gravitational
baryogenesis mechanism in the context of running vacuum models where the background geometry is described by a spatially flat
Friedmann-Lema{\^\i}tre-Robertson-Walker (FLRW) line
element $ds^2 = - dt^2 + a(t)^2 \sum_{i=1}^{3} \left(dx^i\right)^2$, $a (t)$
being the expansion scale factor of the universe.
Recently, a number of cosmological models inspired
from a variable vacuum energy density
have argued that this class of models have better
fit with the observational data in comparison to
the case of a rigid value to  the cosmological
constant \cite{rnv10,rnv11,rnv12,rnv13,coupled4,%
coupled5, coupled6}. In the literature there exists
various proposals of dynamical vacuum models,
but in the present work we follow the approach
of \cite{rnv7,rnv8,rnv9,rg2,Guberina,Shapiro},
where the scaling evolution of the
vacuum energy occurs through a renormalization
group equation for the cosmological constant.

In a spatially flat FLRW universe, the Einstein's equations 
for a perfect fluid plus a dynamic vacuum energy are given by

\begin{eqnarray}
H^2 = \frac{8 \pi G}{3}\,\,\Bigl[  \rho + \rho_{\Lambda}  \Bigr],\label{EFE1}\\
2\dot{H}+ 3 H^2 = -\,8 \pi G\,\, \Bigl[  p +p_{\Lambda} \Bigr],\label{EFE2}
\end{eqnarray}
where the ``dot'' represents the differentiation with respect
to the cosmic time; $H = \dot{a}/a$ is the Hubble rate; $p$, $\rho$
are respectively the pressure and the energy density of
the perfect fluid with equation of state $p= w \rho$,  $w$ being
the equation of state parameter of the perfect fluid,
and $\rho_{\Lambda}= \Lambda(H)/8 \pi G$
is the vacuum energy density associated with the time
dependent $\Lambda (H)$, and $p_{\Lambda}\;(= -\rho_{\Lambda})$
is the pressure of the vacuum energy density.

The continuity equation for the total energy density
$\rho_{tot}= \rho + \rho_{\Lambda}$ is written as,
\begin{equation}\label{cont}
\dot{\rho}+ 3 H (1+w) \rho = -\dot{\rho}_{\Lambda}\, ,
\end{equation}
which can also be obtained by using (\ref{EFE1}) and (\ref{EFE2}).
This conservation equation can also be cast as follows,
\begin{equation}\label{asxet}
\dot{H} + \frac{(1+w)}{2}\, \Bigl[  3\,H^2 - \Lambda (H) \Bigr ] = 0\, ,
\end{equation}
which is the master equation to understand the dynamics
of the universe in the context of RVMs.
Certainly, to go ahead one needs an expression
for $\Lambda (H)$. A series of running vacuum models confronted with 
observational data can be 
found in \cite{Gomez-Valent:2014rxa}. Thus, being motivated from the previous studies, 
we recall the most used RVMs in the literature as follows.

The first RVM that we introduce is given by \cite{Lima:2012mu, Perico:2013mna, Sola2013,SV2015}

\begin{equation}\label{LH}
\mathrm{RVM} \, \mathrm{I}:\,\,\Lambda(H) = \alpha + 3 \, \beta \, H^2 + 3\, \delta \frac{H^{n+2}}{H_I^n},
\end{equation}
where $n$ is any positive integer; $H_I$ is the Hubble
parameter at the inflation scale, $\alpha$, $\beta$, $\delta$ are
free and dimensionless parameters of the model. The parameter
$\alpha$ dominates at low energy (late times), and without any loss
of generality it can be fixed/interpreted as the current value of
the vacuum energy density \footnote{When assessed at present time, the constant $\alpha$ relates to $\alpha = \Lambda_0 - 3 \beta H^2_0$ 
(the third term in the right hand side of (\ref{LH}) can be safely neglected since the scale of $H_I$ is much much larger than $H$ at present). 
Since $\alpha$ is an arbitrary constant, hence, without any loss of generality, we can do
$\alpha = \bar{\Lambda}_0 = \Lambda_0 - 3 \beta H^2_0$. Therefore, the contribution from $\alpha$ 
is despicable during very early times.}. 
The parameter $\delta$ can be absorbed by the
undetermined scale $H_I$. Although the inflation scale is
undetermined, but however, it is well motivated so that one can
consider this to occur on the scales of the order of $10^{16}$ $-$
$10^{19}$ GeV, for example. Thus, in general, the third term in
the right hand side of eqn. (\ref{LH}) is predominant during inflation and despicable on the
other cosmological eras  (radiation, matter and late times), since
$H_I$ is substantially greater than $H$ in other eras. The
parameter $\beta$ can be determined from observations and this
parameter quantifies the dynamic character of the vacuum energy. The
model above is consistent with the general covariance and can exit
from the early inflationary era to the isotropic and homogeneous
radiation dominated phase \cite{Perico:2013mna}. This class of
models can also give an estimation of the ratio of early and present
vacuum energy density \cite{Lima:2012mu}. We note that the model (\ref{LH})
cannot be analytically solved for arbitrary $n$, but however,
the gravitational baryogenesis mechanism can constrain this
$\beta$-parameter under certain conditions, that we will show later.

Another model we shall discuss is \cite{Lima:2012mu, Perico:2013mna, Sola2013,SV2015}

\begin{equation}\label{LH2}
\mathrm{RVM} \, \mathrm{II}: \Lambda (H)= \Lambda_0 + 3 \nu (H^2- H_0^2),
\end{equation}
where $\Lambda_0$ is the current value of the vacuum energy density
and $\nu$ is a dimensionless parameter. The parameter $\nu$ has been
constrained using the observational data which turns out to be of
the order of $\nu \sim \mathcal{O} (10^{-3})$
\cite{rnv10,rnv11,rnv12}. From a recent study
\cite{basil-new2}, the authors have shown that the growth index can
be filed to $\gamma_{\Lambda(H)} = \frac{6+3\nu}{11-12\nu}$,
extending the growth index of the matter perturbations to RVMs. For
a recent analysis of this model including both background and
perturbations analysis (including also full CMB Planck likelihood),
we also refer to \cite{Geng:2017apd}, where $\nu \sim \mathcal{O}
(10^{-4})$ is reported.

In the following sections we shall discuss the implications of these models on the gravitational baryogenesis scenario
and we shall investigate when the results are compatible with the observations.

\section{Gravitational Baryogenesis in Running Vacuum Models}

\label{gravibarmain}

The gravitational baryogenesis mechanism used in \cite{gb1},
makes use of one of the well known Sakharov criteria for baryogenesis,
and it is based on the existence of a
$\mathcal{C}\mathcal{P}$-violating interaction,
which has the following form \cite{gb1},
\begin{equation}\label{baryonassterm}
\frac{1}{M_*^2}\int \mathrm{d}^4x\sqrt{-g}(\partial_{\mu} R) J^{\mu}\, .
\end{equation}
As was stressed in \cite{gb1}, terms of the form
(\ref{baryonassterm}) may occur from higher order interaction originating
in the complete underlying effective theory which
controls the high energy regime. The current term is $J^{\mu}$ and it stands
for the baryon matter current, the parameter $M_*$
represents the cutoff scale of the effective theory while $g$ and $R$
are respectively the trace of the metric tensor $g_{\mu \nu}$
and the Ricci scalar. As was demonstrated in \cite{gb1},
the resulting baryon-to-entropy ratio for the
$\mathcal{C}\mathcal{P}$ violating interaction
of Eq. (\ref{baryonassterm}) is equal to,
\begin{equation}
\label{baryontoentropyrationori}
\frac{n_B}{s}\simeq -\frac{15g_b}{4\pi^2g_*}\frac{\dot{R}}{M_*^2
{T}}\Big{|}_{T_D}\, ,
\end{equation}
where $T_D$ denotes the decoupling temperature, $g_b$ is
the number of the intrinsic degrees of freedom of the
baryons, and $g_*$ is the total number of the degrees of
freedom corresponding to effectively massless particles.

An important assumption that needs to be taken into account is that
the vacuum decays ``adiabatically'', that means the specific entropy
of the massless particles (entropy per massless particle) remains
constant irrespective of the total entropy, which may increase. In
that case, some equilibrium relations are hold not all
\cite{Lima1996}, for instance the energy density versus temperature
($\rho_r \propto T_r^4$), particle number versus temperature  ($n_r
\propto T_r^3$), but $T_r$ does not follow the usual relation $T_r
\propto a(t)^{-1}$. Following the above line of research the
temperature of the massless particle or radiation follows
\cite{thermobas}

\begin{equation}
\label{equilibrium}
\rho_{r}=\frac{\pi^2}{30}g_*\, T_{r}^4\, ,
\end{equation}
where $g_*$ is defined previously.

Let us find an analytic expression for the baryon-to-entropy
ratio by using the running vacuum equations of motion.
The Ricci scalar $R$, in this spacetime can be calculated to be

\begin{equation}\label{einscontr}
R=-8\pi G(1-3w)\rho-32\pi G\rho_{\Lambda}\, ,
\end{equation}
where we assumed the presence of a perfect fluid with equation of state $p=w\rho$, in which $w$ is the equation of state parameter and it is constant.
The expression in eq. (\ref{einscontr}) is very relevant for the calculation of the baryon-to-entropy ratio (\ref{baryontoentropyrationori}),
since the first derivative of the Ricci scalar with respect to the cosmic time appears in eq. (\ref{baryontoentropyrationori}).
By differentiating eq. (\ref{einscontr}) with respect to the cosmic time we obtain,
\begin{equation}\label{firstdercontr}
\dot{R}=-8\pi G(1-3w)\dot{\rho}-32\pi G\dot{\rho}_{\Lambda}\, ,
\end{equation}
and by using the continuity equation (\ref{cont}), one may express the first derivative of the perfect fluid
energy density $\dot{\rho}$ as a function of $\rho$ and $\dot{\rho}_{\Lambda}$. However, in the following sections
we shall focus on the case where the perfect fluid is purely radiation, so in this case $w=1/3$.

\subsubsection{Baryogenesis in the Conformal Case}

There is a sound difference between the Einstein-Hilbert gravitational baryogenesis and the running vacuum gravitational baryogenesis,
for the particular case that the perfect fluid content of the universe is a radiation fluid. In the Einstein-Hilbert case,
the Einstein equations have the following form,
\begin{equation}\label{ricci}
R=-8\pi G (1-3w)\rho\, ,
\end{equation}
so the first derivative of the Ricci scalar is, $\dot{R}=-8\pi G
(1-3w)\dot{\rho}$, and effectively for the a radiation fluid, the
baryon-to-entropy ratio (\ref{baryontoentropyrationori}) is zero. In
the running vacuum case, the first derivative of the Ricci scalar is
given in Eq. (\ref{firstdercontr}), and for $w=1/3$, this becomes,
\begin{equation}
\dot{R}=-32\pi G\dot{\rho}_{\Lambda}\, ,
\end{equation}
which can be further written as,
\begin{equation}
\dot{R}=-4\dot{\Lambda} (H)\, ,
\end{equation}
where we used $\rho_{\Lambda}=\Lambda (H)/8\pi G$. Therefore, even in the conformal case, the resulting baryon-to-entropy ratio is non-zero in the running vacuum case and it is equal to,
\begin{equation}
\label{baryontoentropyrationori1}
\frac{n_B}{s}\simeq \frac{15g_b}{\pi^2g_*}\frac{\dot{\Lambda}(H)}{M_*^2
{T}}\Big{|}_{T_D}\, .
\end{equation}
We remark that if $\Lambda$ is constant, then from eq. (\ref{baryontoentropyrationori1})
one finds that the baryon-to-entropy ratio is zero, which is the usual case.
So, effectively, by having the particular form of the running vacuum function $\Lambda (H)$, one can
easily find the baryon-to-entropy ratio from Eq. (\ref{baryontoentropyrationori1}).
In the following sections we shall concentrate on two particular RVMs (\ref{LH}) and (\ref{LH2})
that have been widely investigated in the literature.

\subsection{Running vacuum model I}

Let us first consider the first RVMs I defined in eq. (\ref{LH}), for which from eq. (\ref{asxet}) becomes

\begin{equation}\label{asxeto1}
\dot{H} + \frac{3\, (1+w)}{2}\, H^2\, \Bigg[ 1- \beta -\frac{\alpha}{3\, H^2}-\delta\, \left(   \frac{H}{H_I} \right)^n  \Bigg] = 0
\end{equation}

Evaluating at the early time, the dynamics is governed by

\begin{equation}\label{opn1}
\dot{H} \simeq - \, \frac{3\, (1+w)}{2}\, H^2\, \Bigg[ 1- \beta -\, \left(   \frac{H}{H_I} \right)^n  \Bigg],
\end{equation}
which shows that for $H \simeq H_I\, \left(1-\beta \right)^{1/n}$ (where we must have $\left(1-\beta \right)> 0$),
the inflationary period can be realized. As mentioned above, the factor $\delta$ was absorbed in $H_I$,
so that it can be fixed to unity without any loss of generality (if the scale of inflation is not precisely known).
This possibility has already been found in \cite{BLS2013}, but our motivation is
to investigate the baryogenesis mechanism in this context. If however, the de Sitter solution is considered,
the predicted baryon-to-entropy ratio in eq. (\ref{baryontoentropyrationori1}) is zero, since in this case the vacuum energy is constant.
Therefore, may be one can think of some time dependent solution for $\Lambda(H)$ that could account for the early inflationary
era and finally at late time it can survive as a solution to the observed accelerated expansion. When evaluated at early times,
the second and third term in the right hand side in eq. (\ref{asxeto1}) can be omitted
\footnote{The constant $\alpha$ is despicable in early times since this factor plays 
like $\alpha=\Lambda_0$, the current value of vaccum energy density which is very small. 
Moreover, the contribution of the term $H^{n}/H_I$, can also be discarded during the radiation era.}

\begin{equation}\label{resdiff}
\dot{H}\simeq -\frac{3\, (1+w) (1-\beta)}{2}H^2\, ,
\end{equation}
which can be analytically solved and the result is,

\begin{equation}\label{htremodel1}
H(t)\simeq \frac{2}{3 (1+w) (1-\beta)t}\, ,
\end{equation}
where without any loss of generality and for simplicity we assume the integration constant to be zero. Within this scenario,
the function $\Lambda (H)$ is approximately equal to $\Lambda (H)\simeq 3\beta H^2$, that is, a pure dynamic vacuum. 
Thus, the resulting baryon-to-entropy ratio in the conformal case, where $w=1/3$, 
can be easily found by using eq. (\ref{baryontoentropyrationori1}). We find,

\begin{equation}\label{nbsmodelianalytf}
\frac{n_B}{s}\simeq \frac{15g_b}{4\pi^2g_*}\frac{6\beta H \dot{H}}{M_*^2
{T}}\Big{|}_{T_D}\, .
\end{equation}

By using Eqs. (\ref{EFE1}) and (\ref{equilibrium}), we can express the decoupling time $t_D$ as a function of the decoupling
temperature $T_D$, resulting to

\begin{equation}\label{dectime}
t_D=\sqrt{\frac{90g_b}{2\pi G}}\,\,\,~\left(\frac{1}{3 (1+w) (1-\beta)\pi T_D}\right)\, ,
\end{equation}

By combining Eqs. (\ref{htremodel1}) and (\ref{nbsmodelianalytf}), the resulting expression for the baryon-to-entropy ratio is given by

\begin{equation}\label{baryonmodel1}
\frac{n_B}{s}\simeq \frac{120 g_b}{g_*M_*^2}\left( \frac{2\pi G}{90g_b}\right )^{3/2}3 (1+w) (1-\beta )\pi^2T_D^2\, .
\end{equation}

Thus, from eq. (\ref{baryonmodel1}) one can conclude that for RVM I,
the calculated baryon-to-entropy ratio should be nonzero since
$\beta \neq 1$. Now, one can find an approximate value of $n_B/s$,
using the values of the free parameters in a reasonable way. To
illustrate this notion, we consider a particular example that will
reflect the nonzero baryon-to-entropy ratio. We consider the cutoff
scale to be $M_*=10^{12}$GeV, in addition that the decoupling
temperature $\mathcal{T}_D$ is equal to $\mathcal{T}_D=M_I=2\times
10^{16}$GeV, with $M_I$  being the upper bound for the fluctuations
constraints on tensor-modes of the inflationary scale, and also that
$g_b\simeq \mathcal{O}(1)$, $g_*\simeq 106$, which is the total
number of effective massless particles at early times that is well
known.

At this point it is worth discussing the values assigned to the
parameters $M_*$, $M_I$ and $T_D$. Recall that $M_*$ is the scale
that the operator appearing in Eq. (\ref{baryonassterm}) becomes
effective, $M_I$ is the inflationary scale, and $T_D$ is the
decoupling temperature. The cutoff scale $M_*$ in a gravitational
context is expected to be much lower than the Planck scale, so we
assumed for example $M_*=10^{12}$GeV, but in principle it can take
alternative values. Now for the purposes of this paper we shall
assume that $M_I\sim \mathcal{T}_D$, which is also the choice made
in Ref. \cite{gb1}. This choice  significantly constrains the
decoupling temperature $\mathcal{T}_D$, since the inflationary scale
is constrained as follows $M_I\leq 3.3\times 10^{16}$GeV \cite{gb1}.
Hence, the choice $\mathcal{T}_D=M_I=2\times 10^{16}$GeV, is
compatible with the above constraint.

Now, with the choices made above, the resulting baryon-to-entropy
ratio becomes very sensible to the parameter $\beta$. For different
values of $\beta$, we can calculate the baryon-to-entropy ratio.
However, for a typical value of $\beta= 0.999$, the resulting
baryon-to-entropy ratio is $n_B/s\simeq 17.9316 \times 10^{-11}$,
which is compatible with the observed value $n_B/s\simeq 9\times
10^{-11}$. On the other hand, for $\beta \gtrsim 1$, the resulting
baryon-to-entropy ratio becomes ``negative'', which means that, the
model with $\beta \gtrsim 1$ predicts an excess of anti-matter over
matter, which is unphysical. Thus, it is evident that the mechanism
of gravitational baryogenesis constrains the RVMs under
consideration. Note that  we are considering the conformal
case, i.e., $w=1/3$. However, one may note that it is possible to
obtain the nonzero $n_B/s$ for some other specific choices of the
free parameters, where we can take different values of $\beta < 1$.
We can note that for $\beta= 0.999$, we have $\Lambda_0 \simeq 3H^2_0$ at late times within this context 
addressed here. This value has the same order of magnitude with the standard case $\Lambda_0 \simeq 2H^2_0$ 
(assuming $\Omega_{\Lambda}=0.70$ and natural units). Therefore, the class RVM I can produce a non-zero and physically
acceptable baryon-to-entropy ratio within this context.

\subsection{Running vacuum model II}

Let us now calculate the predicted baryon-to-entropy ratio for RVM II given by eq. (\ref{LH2}), for the conformal matter fluid case.
Here, the Hubble rate can be found analytically and it reads,

\begin{align}
H  &= \sqrt{\frac{\Omega_{\Lambda0}-\nu}{1-\nu}} \tanh \left[\frac{3 (1+w)}{2}H_p\,\sqrt{(1-\nu)(\Omega_{\Lambda0}-\nu)}\, \left( t-t_p  \right)+ B\right]\, ,
\end{align}
where $H_p= H (t= t_p)$, $B = \tanh^{-1}\left(\sqrt{\frac{1-\nu}{\Omega_{\Lambda0}-\nu}}\, H_p\right)$. Further, one can integrate the above equation
to find the scale factor as,

\begin{equation}
a= a_p\, \Bigg[\frac{\cosh\left(\frac{3}{2}(1+w)H_p\,\sqrt{(1-\nu)(\Omega_{\Lambda0}-\nu)}\, \left( t-t_p  \right)+B\right)}{\cosh B}\Bigg]^{\frac{2}{3 (1+w) H_p (1-\nu)}}\, .
\end{equation}
We can find the leading order expression for the decoupling time as a function of the decoupling temperature, which in this case is,
\begin{equation}\label{dectime1}
t_D=\frac{45 \mathcal{C}_1-4\pi Gg_*T_D^4}{\mathcal{C}_2}\, ,
\end{equation}
where the parameters $\mathcal{C}_1$ and $\mathcal{C}_2$ are defined as

\begin{align}
& \mathcal{C}_1=\frac{(\nu -\Omega_{\Lambda_0}) \tanh \left[B-\frac{3}{2} H_p t_p (1+w) \sqrt{(\nu -1 ) (\nu -\Omega_{\Lambda_0})}\right]^2}{\nu -1 },
\end{align}
\begin{align}
\mathcal{C}_2 =\frac{3 \left((\nu -\Omega_{\Lambda_0}) \left(-H_p (1+w)  \sqrt{(\nu -1 ) (\nu -\Omega_{\Lambda_0})}~ \text{tanh}\left[B-\frac{3}{2} H_p t_p (1+w) \sqrt{(\nu -1 ) (\nu -\Omega_{\Lambda_0})}\right]\right)\right)}{\nu -1 }\nonumber \\ +\frac{\left.3 \left(H_p (1+w)  \sqrt{(\nu -1 ) (\nu -\Omega_{\Lambda_0})}~ \text{tanh}\left[B-\frac{3}{2} H_p t_p (1+w)  \sqrt{(\nu -1 ) (\nu -\Omega_{\Lambda_0})}\right]^3\right)\right)}{\nu -1 }.
\end{align}

Now, for this model
it can be shown that the function $\dot{R}$ is equal to,
\begin{equation}\label{ephe}
\dot{R}\simeq -24\nu \left[\mathcal{A}_1-\frac{\mathcal{A}_2}{\mathcal{C}_2}\left(\mathcal{C}_1-\frac{4\pi G g_*T_D^4}{45} \right) \right]\, ,
\end{equation}
where the parameters $\mathcal{A}_1$ and $\mathcal{A}_2$ can be
calculated as

\begin{align}
 \mathcal{A}_1 & =\frac{3 H_p (1+w)  (\nu -\Omega_{\Lambda_0}) \sqrt{(\nu -1) (\nu -\Omega_{\Lambda_0})}\,\,\, \text{sech}\left[B-\frac{3}{2} H_p t_p (1+w)  \sqrt{(\nu -1 ) (\nu -\Omega_{\Lambda_0})}\right]^2 }{2 (\nu -1 )}\\ \notag & \times \text{tanh}\left[B-\frac{3}{2} H_p t_p (1+w) \sqrt{(\nu -1 ) (\nu -\Omega_{\Lambda_0})}\right],
 \end{align}
 \begin{align}
\mathcal{A}_2 &=\frac{9}{4} \left(H_p^2 (1+w) ^2 (\nu -\Omega_{\Lambda_0})^2 \left(-~\text{sech}\left[B-\frac{3}{2} H_p t_p (1+w) \sqrt{(\nu -1 ) (\nu -\Omega_{\Lambda_0})}\right]^2\right.\right.\\ \nonumber &
\left.\left.+3~ \text{sech}\left[B-\frac{3}{2} H_p t_p (1+w)  \sqrt{(\nu -1 ) (\nu -\Omega_{\Lambda_0})}\right]^2 \text{tanh}\left[B-\frac{3}{2} H_p t_p (1+w) \sqrt{(\nu-1 ) (\nu -\Omega_{\Lambda_0})}\right]^2\right)\right)\, .
\end{align}

Therefore, the resulting baryon-to-entropy ratio for RVM II is given by
\begin{equation}\label{baryonmodel2analyticexpre2}
\frac{n_B}{s}\simeq \frac{90 \nu g_g}{\pi g_*M_*^2T_D}\left[\mathcal{A}_1-\frac{\mathcal{A}_2}{\mathcal{C}_2}\left(\mathcal{C}_1-\frac{4\pi G g_*T_D^4}{45} \right) \right]\, .
\end{equation}

In this case we assume the cutoff scale
$M_*=10^{12}$GeV, in addition to that,  the decoupling temperature $T_D$ to $\mathcal{T}_D=M_I=0.23\times 10^{13}$ GeV,
$g_b\simeq \mathcal{O}(1)$, $t_p\simeq 10^{-20}$sec, $g_*\simeq 106$, $H_p/\Omega_{\Lambda_0}=10^{-6}$, and $\nu =0.001$
\footnote{This value is fixed from the results in \cite{rnv10,rnv11,rnv12}.}.

Notice that, as in the previous model, we assume that
$M_*=10^{12}$GeV, however, in this case we made the choice
$\mathcal{T}_D=M_I=0.23\times 10^{13}$ GeV, which is compatible with
the observational constraint on the inflationary scale $M_I\leq
3.3\times 10^{16}$GeV \cite{gb1}. 

By using the above mentioned values for the parameters, we find that
the baryon-to-entropy ratio is approximately equal to $n_B/s\simeq
7.29947 \times 10^{-11}$, which is again compatible with the
observed value of $n_B/s\simeq 9\times 10^{-11}$.

Therefore, for the RVM II model too, even in the conformal case, the
predicted baryon-to-entropy ratio is non-zero. In principle, the
function $\Lambda (H)$ crucially affects the predicted
baryon-to-entropy ratio. Thus, it is possible that the gravitational
baryogenesis mechanism may pose some constraints on the general class
of models where the vacuum energy is variable.

\section{Conclusions}

In this paper we investigated the gravitational baryogenesis mechanism in the context of running vacuum models,
regardless if these scenarios can produce a viable baryon-to-entropy ratio. We found that there is a sound difference between
the Einstein-Hilbert baryon-to-entropy ratio and the one corresponding the scenarios with running vacuum energy. Particularly,
when the universe is filled with conformal matter fluids, the baryon-to-entropy ratio is zero in the Einstein-Hilbert case,
whereas in the running vacuum scenario the predicted baryon-to-entropy ratio is non-zero. We calculated in some detail the baryon-to-entropy
ratio for two well known RVM models and we demonstrated that the resulting picture is compatible with the observational data.
From both the models, we came up with the conclusion that in the case of RVMs the resulting baryon-to-entropy ratio
is non-zero even in the presence of conformal matter. We also observed that the gravitational baryogenesis mechanism can be used to constrain
the RVM models. In general, a nonzero ratio of baryon-to-entropy is sensitive to the values of $\beta$ (RVM I) and $\nu$ (RVM II).
Thus, it is clear that cosmological models with running vacuum energy carry a
possible explanation to the non-zero baryon-to-entropy ratio in our universe and conversely the observed value of the baryon-to-entropy ratio
can also be used to constrain the RVM models as well.
Although the current work focuses on two well known models of running vacuum energy, but it would be interesting to study the cosmological
scenarios driven by another models in order to measure the baryogenesis effect which could potentially impose some constraints on these
phenomenological scenario.

Finally, we close the work drawing a possible connection of this scenario with  the extended gravity theories. It is well known that the physics of the universe at early times constitutes a great laboratory to explore physics beyond the standard cosmological model. For instance, torsion in $f(T)$
theories of gravity can be constrained by big bang nucleosynthesis \cite{Capozziello:2017bxm},
modifications on the evolution of the thermal relic particles (particles like WIMPs (weakly interactive massive particles)) within $f(R)$ gravity context are also taken into account
in \cite{Capozziello:2015ama}. Thus, it is worth noting that the nonzero ratio of baryon-to-entropy of the universe could be a potential quantity to constrain extended theories of gravity too.

\section*{Acknowledgments}

The authors gratefully acknowledge the referees for their comments to improve the work. The authors also thank Prof. J. A. S. Lima for some useful comments. This work is supported respectively by the ``Min. of Education and Science of Russia'' (V.K.O) and ``SERB-NPDF (File No: PDF/2015/000640)'', Govt. of India  (S.P).

\end{document}